# The Search for Dark Matter, Einstein's Cosmology and MOND

David B. Cline

Astrophysics Division, Department of Physics & Astronomy
University of California, Los Angeles, CA 90095 USA
dcline@physics.ucla.edu

Abstract

The discovery of dark matter particles would conclusively reject the MOND theory. MOND may violate Einstein's Strong Equivalence principle. However, as we show, there is already evidence that MOND is likely not required. MOND was invented to explain the rotation velocities of stars far into the galactic halos. Dark matter also explains this same effect.
These both use a gravity probe of the $1/R^2$ law. We show that non gravity probes determine the same value for the amount of dark matter that does not involve modifications of gravity. Using Occam's Razor this coincidence is best explained by the existence of dark matter.

Introduction

The search for dark matter is of great importance in the context of general theory of relativity. A brief history of dark matter/dark energy is given in Table 1.
There are suggested modifications to take into account the effects attributed to dark matter. The most noticeable is the MOND theory of Milgram [1]. Some say that this theory violates Einstein's Strong Equivalence principle---and therefore general relativity. Thus it is important to determine its correctness.
The MOND theory was invented to explain the star rotation curves for galaxies that are normally used to indicate the existence of cold dark matter. MOND assumes that gravity is modified beyond a gravitational acceleration of $a_0 \sim 10^{-8}$ cm/Sec$^2$ [1,2]

For acceleration greater than this we get the normal law:

$$a_N = \frac{MG}{R^2}$$

for acceleration less than $a_0$ we get

$$a = \sqrt{a_0 a_N} = \sqrt{a_0 MG}\,\frac{1}{R}$$

The 1/R behavior then provides a good fit to the rotation curves of galaxies at large r.
There are three more (at least) ways to test MOND [3]:
a) Study of clusters in x-rays and weak lensing to determine whether dark matter is different from Baryonic matter

b) Study of CMBR and Baryon oscillations that come from Z ~ 1000 – the surface of last scatter – two independent ways to measure $\Omega_{DM}$
c) Direct detection of dark matter particles (reviewed here)

**Table 1.**
**Brief history of the evidence for dark matter and dark energy**

~1933        F. Zwicky observes fast galaxies in comet clusters; suggests mixing mass in cluster is the cause

~1960s       Astronomers realize that galaxies have fast moving stars in halo; suggest dark matter is the cause

~1980        Suggestions for MOND to explain radiation curves by modifying Newtonian gravity [1][2]

~1998        Experiments on SN1A reports at dark matter meeting; indication of accelerating universe; dark energy is suggested cause

~2003        WMAP sata strongly supports both dark matter and dark energy components of the universe

~2005        SDSS observes Baryon acoustic oscillations; provides additional proof for dark matter*

* Several independent measurements of $\Omega_{DM}$ suggest a single origin: cold dark matter

**2. Test of MOND with Gravity Lensing and X-ray Maps**

The MOND theory describes galaxies as being made of only Baryonic matter and with a modified Newtonian force law to explain high velocities in the halo. Any experiment that

1) finds evidence for dark matter with non gravitational probes or
2) finds dark matter displaced from Baryonic matter

will constrain MOND. Two such experiments have been published. We show in Fig. 1 and 2 the results of these experiments [4][5].
    Figure 1 shows the predicted temperature profile for the x-rays from a cluster and the measured profile. In this version of MOND there is a poor fit to the data. In Fig. 2 is shown the results of the study of interacting superclusters. The authors state that the location of the bulk of this mass (determined by weak lensing) is in a different place from the location of the visible (Baryonic matter). Since in the MOND theory there is only Baryonic matter the authors state that this indicates the existence of dark matter.

3. Baryonic acoustic oscillations and evidence for dark matter compared to CMBR evidence

Recently the SDSS collaboration has carried our an impressive study of galactic correlations showing a "Baryonic Oscillation" peak in the spectrum that derives its origin from the early universe. Table 2 lists some of the properties. The measurement of dark matter

$$\Omega_{DM} = 0.275 \pm 0.025$$

for a flat, $\omega = -1$ universe is in impressive agreement with the results for the CMBR from WMAP. Furthermore the authors state this effect is a direct proof of the existence of dark matter. Fig. 3 shows the Baryon oscillation structure.

**Table 2.**
**Testing MOND with Acoustic Oscillations in the Baryon Density**

D. Eisenstein et al, SDSS team (Astro-PH/0501171)

"The recent observation by the SDSS team of an <u>Acoustic Peak</u> in the LSS at a scale of about 100MPC strongly supports the CDM origin of dark matter! One can show that this proves CDM must exist in the redshift ranges:

$$0 < Z < 0.35$$
$$0.35 < Z > 1000$$

for the growth structure.

<u>It will be nearly impossible for MOND to produce such an effect.</u>"

It is hard to see how the MOND theory could reproduce such an effect including the precision value for $\Omega_{DM}$ from these different measurements.
In fact as more experiments measure $\Omega_{DM}$ with different methods the good agreement of the values makes other explanations even less plausible.

**4. The Direct Search for Dark Matter with a Series of Liquid Argon Xenon Detectors**

Starting in 1995 the UCLA – Torino group carried out an R&D program to study the properties of liquid Xenon for dark matter detection [7][8]. One result of this program was the invention of the 2-phase detector that provides powerful discrimination against various kinds of background. The ZEPLIN program of the UKDMC-UCLA-TAMU and other institutions has come particularly from these early results. Currently ZEPLIN II is underground at the Boulby site starting to take data. This is the largest detector in the world with direct discrimination of background [8].

In Figure 4 we show two generations of the liquid Xenon. This detector started with the 1 kg test detector constructed at CERN in 1995, the ZEPLIN I detector (now finished data taking) and the ZEPLIN II detector, now underground at the Boulby Laboratory in the UK, and the proposed ZEPLIN IV detector for the SNOLAB Laboratory in Canada [9]. Figure 5 shos the reach of these experiments superimposed on the theoretical calculations of Pran Nath in the SUGRA model. In this context [10] if dark matter follows the SUGRA model it will be discovered in the next five years.

The direct observation of dark matter would be the final test of MOND.

I wish to thank H. Wang for discussions and the organizers of this excellent meeting.

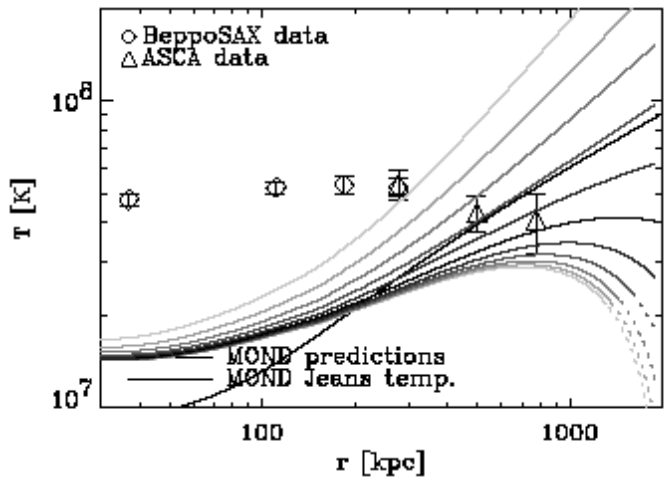

Fig. 1. Temperature profile from x-rays for a galactic cluster from Ref. 4.

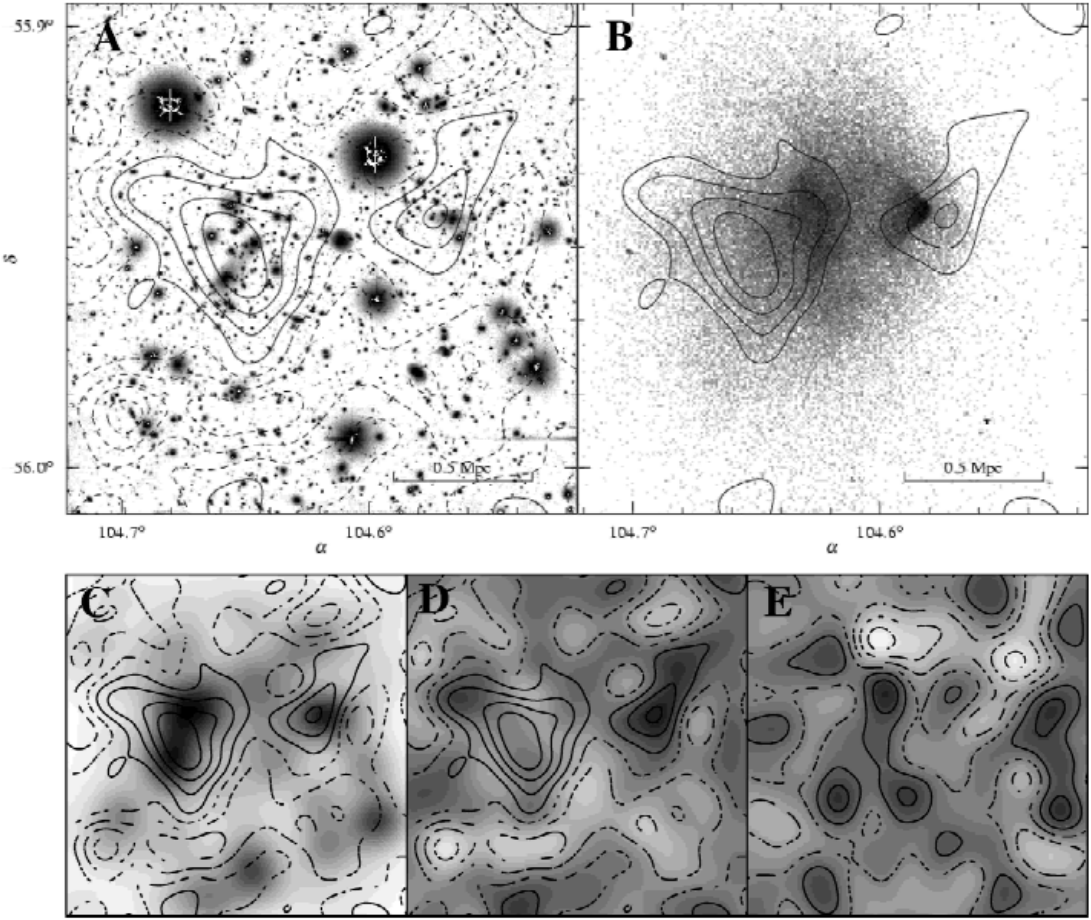

Fig. 2. Images of weak lensing and x-ray studies from Ref. 5.

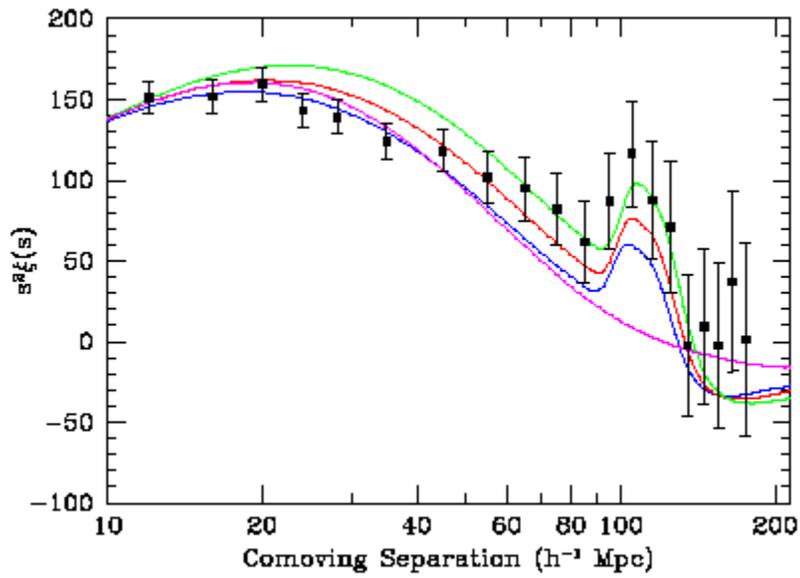

Fig. 3. Evidence for Baryon acoustic oscillations signal from Ref. 6.

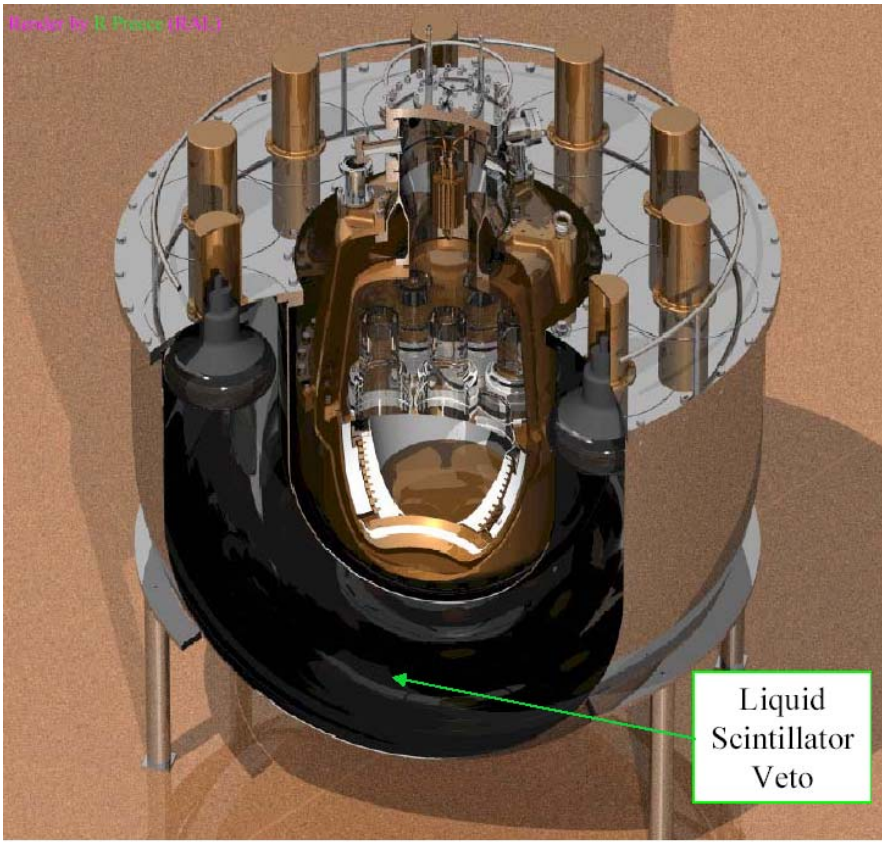

Fig. 4a. Schematic of ZEPLIN IV Detector

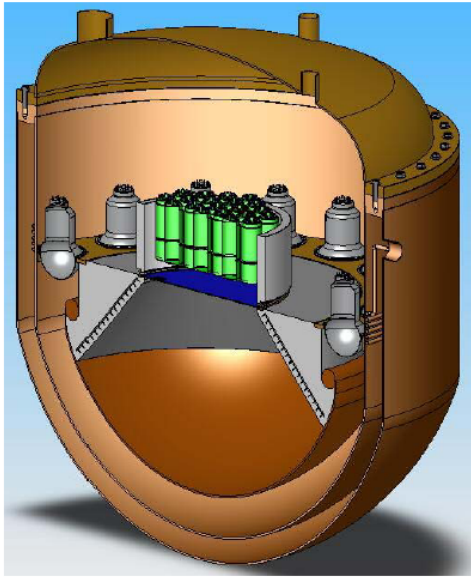
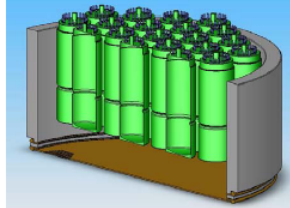
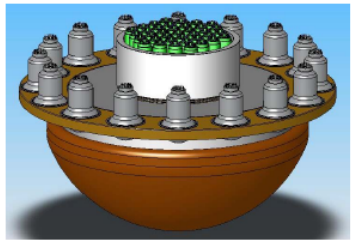

Fig. 4b. Schematic of one version of ZEPLIN IV/MAX (H. Wang, private communication).

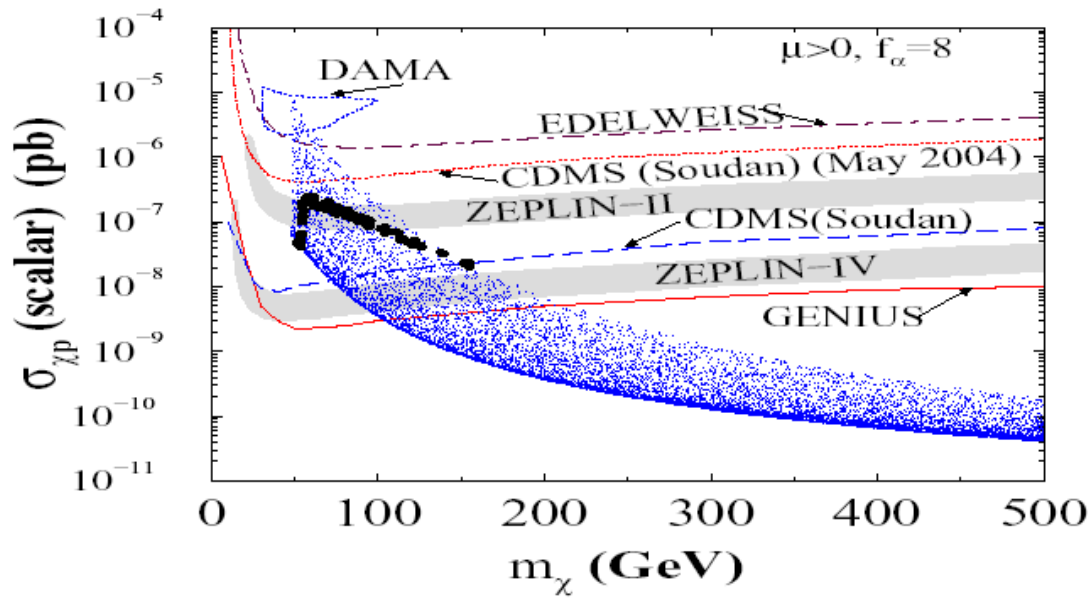

Fig. 5. Expectations for SUSY WIMP cross-sections by Pran Nath and colleagues.